\documentstyle[12pt,twoside,fleqnR,espcrcR,epsfig]{article}

\newcommand{\AmS}{{\protect\the\textfont2
  A\kern-.1667em\lower.5ex\hbox{M}\kern-.125emS}}

\newcommand{\case}[2]{\mbox{\footnotesize $\displaystyle \frac{#1}{#2}$}}
\newcommand{\lsim}{\mathrel{\rlap{\lower4pt\hbox{\hskip0pt$\sim$}}
\raise2pt\hbox{$<$}}}
\newcommand{\gsim}{\mathrel{\rlap{\lower4pt\hbox{\hskip0pt$\sim$}}
\raise2pt\hbox{$>$}}}

\def\Journal#1#2#3#4{{#1} {\bf #2} (#3) #4}
\def\NPB{{\em Nucl. Phys.} B}
\def\PLB{{\em Phys. Lett.}  B}
\def\PRL{\em Phys. Rev. Lett.}
\def\PRC{{\em Phys. Rev.} C}
\def\PRD{{\em Phys. Rev.} D}

\title{Deconfinement and Hadron Properties at \\ Extremes of Temperature and
        Density\thanks{A combined summary of two presentations, one by each
        author.  This work was supported by the US Department of Energy,
        Nuclear Physics Division, under contract number W-31-109-ENG-38; the
        National Science Foundation under grant no. INT-9603385; Deutscher
        Akademischer Austauschdienst; and benefited from the resources of the
        National Energy Research Scientific Computing Center.}}

\author{David Blaschke\address{Fachbereich Physik, Universit\"at Rostock,
                D-18051 Rostock, Germany}        
        and 
        Craig D. Roberts\address{Physics Division,
        Argonne National Laboratory, Argonne IL 60439-4843, USA}}

\begin{document}
% typeset front matter
\maketitle

\begin{abstract}
After introducing essential, qualitative concepts and results, we discuss the
application of Dyson-Schwinger equations to QCD at finite $T$ and $\mu$.  We
summarise the calculation of the critical exponents of two-light-flavour QCD
using the chiral and thermal susceptibilities; and an algebraic model that
elucidates the origin of an anticorrelation between the $\mu$- and
$T$-dependence of a range of meson properties.  That model also provides an
algebraic understanding of why the finite-$T$ behaviour of bulk thermodynamic
properties is mirrored in their $\mu$-dependence, and why meson masses
decrease with $\mu$ even though $f_\pi$ and $-\langle \bar q q\rangle$
increase.  The possibility of diquark condensation is canvassed.  Its
realisation is uncertain because it is contingent upon an assumption about
the quark-quark scattering kernel that is demonstrably false in some
applications; e.g., it predicts the existence of coloured diquarks in the
strong interaction spectrum, which are not observed.
\end{abstract}

\section{DYSON-SCHWINGER EQUATIONS}
The Dyson-Schwinger equations (DSEs) provide a Poincar\'e invariant,
continuum approach to solving quantum field theories.  There are many
familiar examples, among them: the gap equation in superconductivity; and the
Bethe-Salpeter equation (BSE) and covariant Fadde'ev equation, which describe
relativistic 2- and 3-body bound states.  The DSEs are a system of coupled
integral equations and a truncation is necessary to obtain a tractable
problem.  The simplest truncation scheme is a weak-coupling expansion, which
generates every diagram in perturbation theory.  Hence, in the intelligent
application of DSEs to QCD, there is always a tight constraint on the
ultraviolet behaviour.  That is crucial in extrapolating into the infrared,
and in developing uniformly valid, efficacious, symmetry-preserving
truncations.

The task of development is not a purely numerical one, and neither is it
always obviously systematic.  For some, this last point diminishes the appeal
of the approach.  However, with growing community involvement and interest,
the qualitatively robust results and intuitive understanding that the DSEs
can provide is becoming clear.  Indeed, those familiar with the application
of DSEs in the late-70s and early-80s might be surprised with the progress
that has been made.

The DSEs have been used extensively$\,$\cite{rw94} in developing an
understanding of confinement and dynamical chiral symmetry breaking (DCSB),
and their wide application$\,$\cite{pctrev} to the description of hadron
properties in terms of their quark and gluon constituents is built upon that
success.  In understanding and unifying these phenomena, the DSEs point to
the key role played by the necessary, momentum-dependent dressing of the
elementary propagators and vertices in QCD.

\subsection{Gluon propagator}
\begin{figure}[t]
\centering{\ \epsfig{figure=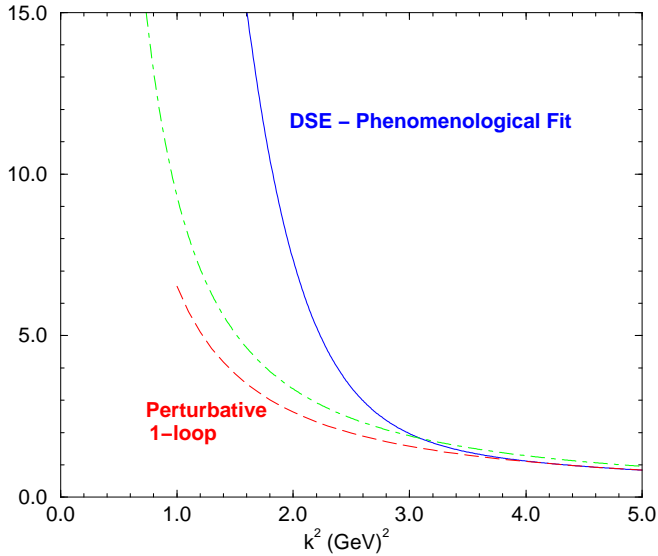,height=7.5cm}}\vspace*{-2.0em}
\caption{${\cal G}(k^2)/k^2$ from a solution$\,$\protect\cite{bp89} of the
gluon DSE (dash-dot line) compared with the one-loop perturbative result
(dashed line) and a fit (solid line) obtained following the method of
Ref.$\,$\protect\cite{mr97}; i.e., requiring that the gluon propagator lead, via
the quark DSE, to a good description of a range of hadron observables.
\label{gluonpic}}\vspace*{-1.0em}
\end{figure}

Important in the application of DSEs is the gluon propagator, which at $T=0$
has the form:
\begin{eqnarray}
g^2 D_{\mu\nu}(k)& = &
\left(\delta_{\mu\nu} - \frac{k_\mu k_\nu}{k^2}\right)
        \frac{{\cal G}(k^2)}{k^2}\,,\; 
        {\cal G}(k^2):= \frac{g^2}{[1+\Pi(k^2)]} \,,
\end{eqnarray}
where $\Pi(k^2)$ is the vacuum polarisation that contains all the dynamical
information about gluon propagation.  Studies of the gluon DSE have been
reported by many authors$\,$\cite{rw94} with the conclusion that, if the
ghost-loop is not significant, then the charge-antiscreening 3-gluon vertex
dominates and, relative to the free gauge boson propagator, the dressed gluon
propagator is significantly enhanced in the vicinity of $k^2=0$.  The
enhancement persists to $k^2 \sim 1$-$2\,$GeV$^2$, where a perturbative
analysis becomes quantitatively reliable.  In the neighbourhood of $k^2=0$
the enhancement can be represented$\,$\cite{bp89} as a regularisation of
$1/k^4$ as a distribution, and this behaviour is illustrated in
Fig.$\,$\ref{gluonpic}.
A dressed-gluon propagator of this type generates DCSB and confinement {\it
without} fine-tuning.  (We identify confinement as the absence of a Lehmann
representation for coloured propagators, with the obvious analogue for other
coloured n-point functions.)

\subsection{Quark propagator}
\begin{figure}[t]
\vspace*{-1.5em}

\centering{\ 
\epsfig{figure=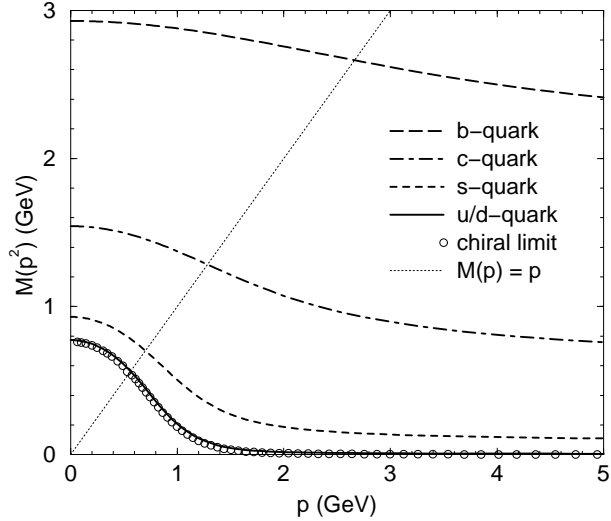,height=7.90cm}}\vspace*{-3.0em}
\caption{$M(p^2):= B(p^2)/A(p^2)$  obtained in solving the quark DSE.  The
solution of $M^2(p^2)=p^2$ defines $M^E$, the Euclidean constituent-quark
mass.
\label{plotMpp}}\vspace*{-1.0em}
\end{figure}

At $T=0=\mu$ in a covariant gauge the dressed-quark propagator has the form
\begin{equation}
\label{Sp}
S(p)  :=  \frac{1}{i\gamma\cdot p + \Sigma(p)}
 :=  \frac{1}{i\gamma\cdot p\,A(p^2) + B(p^2)}  \,.
\end{equation}
$\Sigma(p)$ is the renormalised dressed-quark self energy, which satisfies
\begin{equation}
\label{gendse}
\Sigma(p)  =  ( Z_2 -1)\, i\gamma\cdot p + Z_4\,m^\zeta
+\, Z_1\, \int^\Lambda_q \,
g^2 D_{\mu\nu}(p-q) \frac{\lambda^a}{2}\gamma_\mu S(q)
\Gamma^a_\nu(q,p), 
\end{equation}
where $\Gamma^a_\nu(q;p)$ is the dressed-quark-gluon vertex, $m^\zeta$ is the
current-quark mass, $\zeta$ is the renormalisation point, and $\int^\Lambda_q
:= \int^\Lambda d^4 q/(2\pi)^4$ represents mnemonically a {\em
translationally-invariant} regularisation of the integral, with $\Lambda$ the
regularisation mass-scale.  Using ${\cal G}(k^2)$ similar to that depicted in
Fig.$\,$\ref{gluonpic} and current-quark masses corresponding to
\begin{equation}
\label{monegev}
\begin{array}{llll}
m_{u/d}^{1\,{\rm GeV}} &
m_s^{1\,{\rm GeV}}     & 
m_c^{1\,{\rm GeV}}     & 
m_b^{1\,{\rm GeV}}     \\
 6.6\, {\rm MeV} & 
 140\,{\rm MeV}  & 
 1.0\,{\rm GeV}  & 
 3.4\,{\rm GeV}
\end{array}
\end{equation}
one obtains$\,$\cite{mr98} the dressed-quark mass function depicted in
Fig.$\,$\ref{plotMpp}.

For light quarks ($u$, $d$ and $s$) there are two distinct domains:
perturbative and nonperturbative.  In the perturbative domain the magnitude
of $M(p^2)$ is governed by the the current-quark mass. For $p^2< 1\,$GeV$^2$
the mass-function rises sharply.  This is the nonperturbative domain where
the magnitude of $M(p^2)$ is determined by the DCSB mechanism; i.e., the
enhancement in the dressed-gluon propagator.  This emphasises that DCSB is
more than just a nonzero value of the quark condensate in the chiral limit!

For a given flavour, the ratio ${\cal L}_f:=M^E_f/m_f^\zeta$ is a single,
quantitative measure of the importance of the DCSB mechanism in modifying
that quark's propagation characteristics.  As illustrated in
Eq.$\,$(\ref{Mmratio}),
\begin{equation}
\label{Mmratio}
\begin{array}{l|c|c|c|c|c}
\mbox{\sf flavour} 
        &   u/d  &   s   &  c  &  b  &  t \\\hline
 \frac{M^E}{m^{\zeta\sim 20\,{\rm GeV}}}
       &  150   &    10      &  2.3 &  1.4 & \to 1
\end{array}
\end{equation}
this ratio provides for a natural classification of quarks as either light or
heavy.  For light-quarks ${\cal L}_f$ is characteristically $10$-$100$ while
for heavy-quarks it is only $1$-$2$.  The values of ${\cal L}_f$ signal the
existence of a characteristic DCSB mass-scale: $M_\chi$. At $p^2>0$ the
propagation characteristics of a flavour with $m_f^\zeta< M_\chi$ are altered
significantly by the DCSB mechanism, while for flavours with $m_f^\zeta\gg
M_\chi$ it is irrelevant, and explicit chiral symmetry breaking dominates.
It is apparent from Eq.$\,$(\ref{Mmratio}) that $M_\chi \sim 0.2\,$GeV$\,\sim
\Lambda_{\rm QCD}$.  This forms a basis for many simplifications in the study
of heavy-meson observables.\cite{misha}

\subsection{Hadrons are bound states}
The properties of hadrons can be understood by studying covariant bound state
equations: the Bethe-Salpeter equation (BSE) for mesons and the covariant
Fadde'ev equation for baryons.  The mesons have been studied most extensively
and their internal structure is described by a Bethe-Salpeter amplitude
obtained as a solution of
\begin{eqnarray}
\label{genbse}
\left[\Gamma_H(k;P)\right]_{tu} &= & 
\int^\Lambda_q  \,
[\chi_H(q;P)]_{sr} \,K^{rs}_{tu}(q,k;P)\,,
\end{eqnarray}
where $\chi_H(q;P) := {\cal S}(q_+) \Gamma_H(q;P) {\cal S}(q_-)$; ${\cal
S}(q) := {\rm diag}(S_u(q),S_d(q),S_s(q), \ldots)$; $q_+=q + \eta_P\, P$,
$q_-=q - (1-\eta_P)\, P$, with $P$ the total momentum of the bound state and
observables independent of $\eta_P$; and $r$,\ldots,$u$ represent colour-,
Dirac- and flavour-matrix indices.  

In Eq.$\,$(\ref{genbse}), $K$ is the renormalised, fully-amputated,
quark-antiquark scattering kernel and important in the successful application
of DSEs is that it has a systematic skeleton expansion in terms of the
elementary, dressed-particle Schwinger functions; e.g., the dressed-quark and
-gluon propagators.
\begin{figure}[t]
  \centering{\ \hspace*{-2.5cm}\epsfig{figure=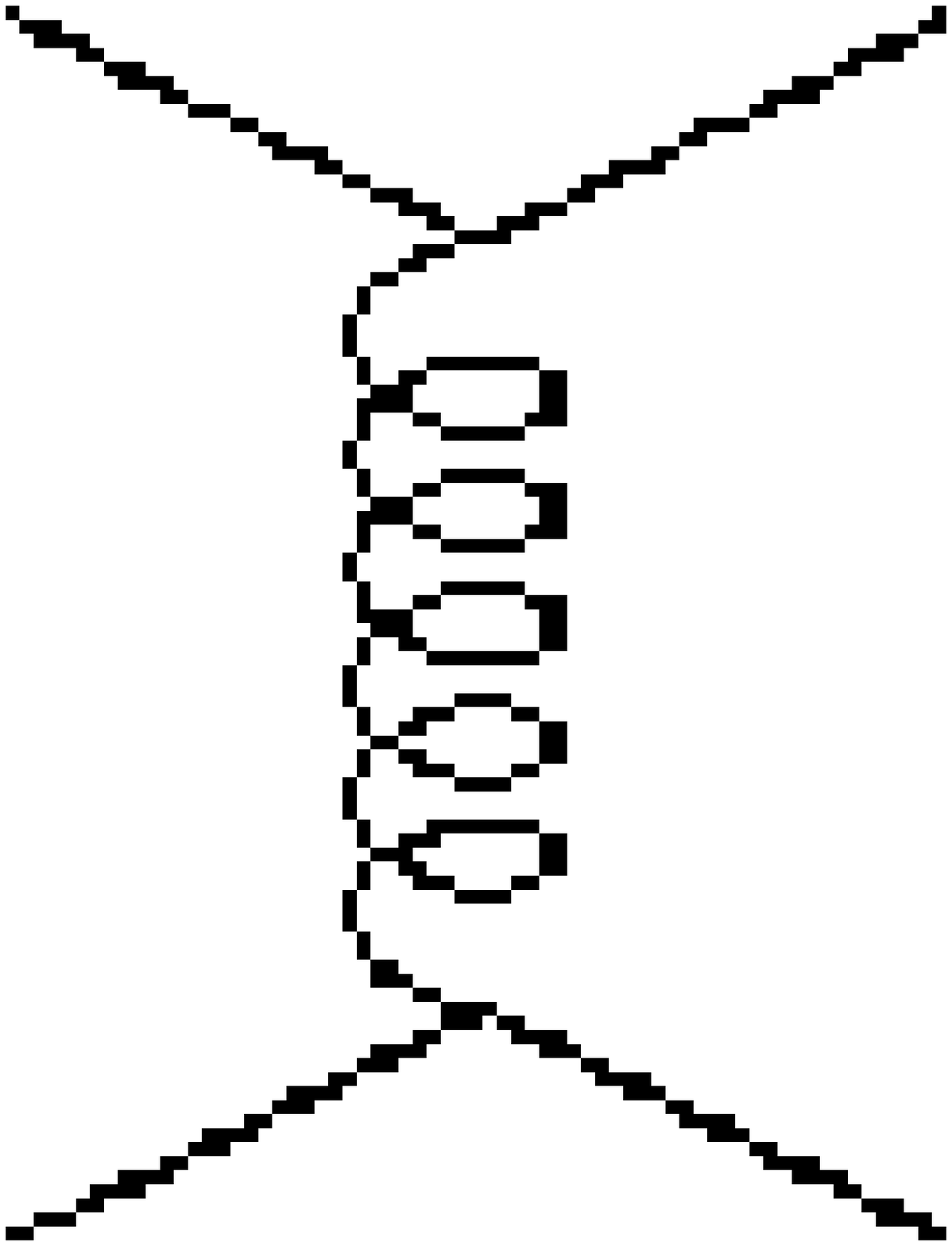,height=2.5cm} 
        \vspace*{-17mm} 

        \hspace*{15mm} $\longleftarrow$ { (1) -- Ladder}\vspace*{4mm} 

        \hspace*{55mm} { (2) -- Beyond Ladder}

        \hspace*{16mm}$\swarrow$

        \hspace*{0.5cm}\epsfig{figure=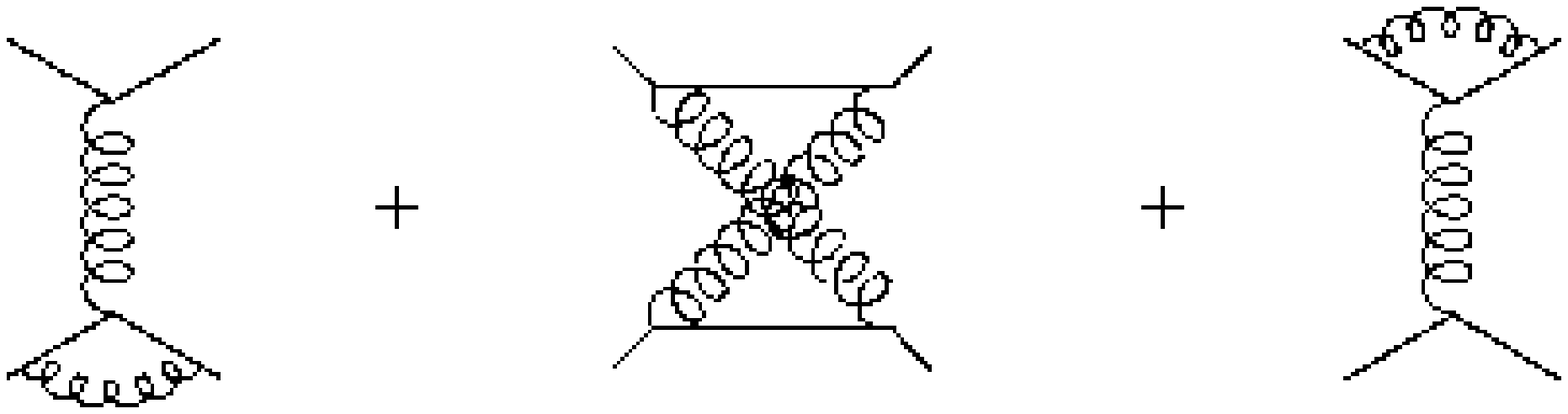,height=2.5cm} }\vspace*{-2.0em}
\caption{First two orders in a systematic
expansion$\,$\protect\cite{bender96} of the quark-antiquark scattering
kernel.  In this expansion, the propagators are dressed but the vertices are
bare.
\label{skeleton}}\vspace*{-1.0em}
\end{figure}
The particular expansion depicted in Fig.$\,$\ref{skeleton}, with its analogue
for the kernel in the quark DSE, provides a means of constructing a kernel
that, order-by-order in the number of vertices, ensures the preservation of
vector and axial-vector Ward-Takahashi identities and hence Goldstone's
theorem.   

\subsection{Phenomenological applications}
\begin{figure}[t]
\vspace*{-1.0em}

\centering{\ 
\epsfig{figure=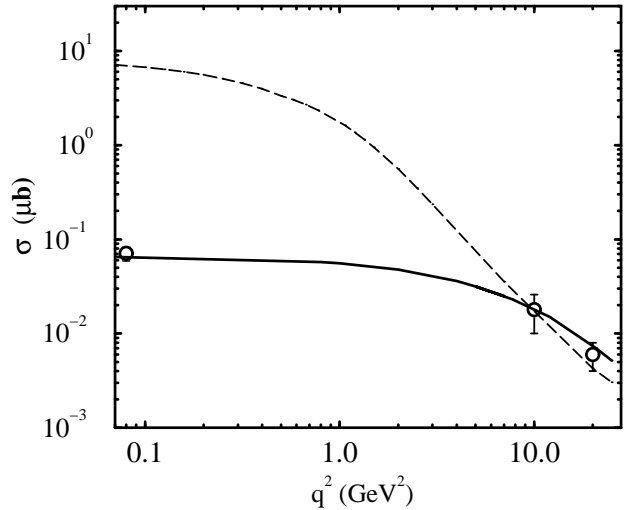,height=7.5cm}}\vspace*{-2.0em}
\caption{$\psi$-meson electroproduction cross section: solid line; the dashed
line is the $\rho$-meson cross section for comparison.  The data are from
Refs.$\,$\protect\cite{derrickcaid}.
\label{psiEP}}\vspace*{-1.0em}
\end{figure}

Some of the many applications of DSEs to the calculation of hadron
observables at $T=0=\mu$ are summarised in Refs.$\,$\cite{pctrev} so here we make
only two brief observations.  Using the DSEs one obtains a mass formula for
pseudoscalar mesons that unifies the small and large current-quark mass
regimes.  For mesons composed of quarks with small current-quark masses the
formula yields$\,$\cite{mr97} the Gell-Mann--Oakes--Renner relation.
However, when the current-quark mass of one or both of the constituents is
large, the formula predicts$\,$\cite{mr98A} that the meson mass increases
linearly with the current-quark mass.  In model studies$\,$\cite{mr98} the
linear increase is dominant for masses as low as that of the $s$-quark.  The
approach has also been extensively employed in the study of scattering
processes, and one example is the calculation$\,$\cite{pichowsky} of the
cross section for the diffractive electroproduction of vector mesons.  This
application makes the striking prediction, confirmed by recent data, that
although two-orders of magnitude smaller than the $\rho$-meson cross section
at the photoproduction point, the $\psi$-meson cross section is equal to that
of the $\rho$-meson at $Q^2=15\,$GeV$^2$, Fig.$\,$\ref{psiEP}.

\section{FINITE TEMPERATURE AND CHEMICAL POTENTIAL}    
The contemporary application of DSEs at finite temperature and chemical
potential is a straightforward extension of the $T=0=\mu$ studies.  The
direct approach is to develop a finite-$T$ extension of {\it Ans\"atze} for
the dressed-gluon propagator and then solve the quark DSE.  Having the
dressed-quark and -gluon propagators, the response of bound states to
increases in $T$ and $\mu$ can be calculated.  As a nonperturbative approach
that allows the simultaneous study of confinement and DCSB, the DSEs have a
significant overlap with lattice simulations: each quantity that can be
estimated using lattice simulations can also be calculated using DSEs.  That
means they can be used to check the lattice simulations, and importantly,
that lattice simulations can be used to constrain their model-dependent
aspects.  Once agreement is obtained on the common domain, the DSEs can be
used to explore phenomena presently inaccessible to lattice simulations.

The renormalised dressed-quark propagator at finite-$(T,\mu)$ has the form
\begin{eqnarray}
\label{genformS}
S^{-1}(\vec{p},\tilde\omega_k)  & = & 
        i\vec{\gamma}\cdot \vec{p}\,A(\vec{p},\tilde\omega_k)
+i\gamma_4\,\tilde\omega_k C(\vec{p},\tilde\omega_k)+
B(\vec{p},\tilde\omega_k)
\end{eqnarray}
where $\tilde\omega_k := \omega_k + i \mu $ with $\omega_k= (2 k + 1)\,\pi
\,T$, $k\in {\rm Z}\!\!\!{\rm Z}$, and satisfies the DSE 
\begin{equation}
\label{qDSE}
S^{-1}(\vec{p},\tilde\omega_k) = Z_2^A \,i\vec{\gamma}\cdot \vec{p} + Z_2 \,
(i\gamma_4\,\tilde\omega_k + m_{\rm bm})\, 
        + \Sigma^\prime(\vec{p},\tilde\omega_k)\,;
\end{equation}
where the regularised self energy is
\begin{eqnarray}
\Sigma^\prime(\vec{p},\tilde\omega_k) & = & i\vec{\gamma}\cdot \vec{p}
\,\Sigma_A^\prime(\vec{p},\tilde\omega_k) + i\gamma_4\,\tilde\omega_k
\,\Sigma_C^\prime(\vec{p},\tilde\omega_k) +
\Sigma_B^\prime(\vec{p},\tilde\omega_k)\; ,
\end{eqnarray}
\begin{equation}
\Sigma_{\cal F}^\prime(\vec{p},\tilde\omega_k)  = 
\int_{l,q}^{\bar\Lambda}\, \case{4}{3}\,g^2\,
D_{\mu\nu}(\vec{p}-\vec{q},\tilde\omega_k-\tilde\omega_l)\case{1}{4} {\rm
tr}\left[{\cal P}_{\cal F} \gamma_\mu
S(\vec{q},\tilde\omega_l)
\Gamma_\nu(\vec{q},\tilde\omega_l;\vec{p},\tilde\omega_k)\right]\,,
\label{regself}
\end{equation}
$\int_{l,q}^{\bar\Lambda}:=\, T
\,\sum_{l=-\infty}^\infty\,\int^{\bar\Lambda}\frac{d^3q}{(2\pi)^3}$ and
${\cal P}_A:= -(Z_1^A/p^2)i\gamma\cdot p$, ${\cal P}_B:= Z_1 $, ${\cal P}_C:=
-(Z_1/\tilde\omega_k)i\gamma_4$.  The complex scalar functions:
$A(\vec{p},\tilde\omega_k)$, $B(\vec{p},\tilde\omega_k)$ and
$C(\vec{p},\tilde\omega_k)$ satisfy:
$ {\cal F}(\vec{p},\tilde\omega_k)^\ast = {\cal
F}(\vec{p},\tilde\omega_{-k-1})\,, $
${\cal F}=A,B,C$, and although not explicitly indicated they are functions
only of $|\vec{p}|^2$ and $\tilde\omega_k^2$.  The dependence of these
functions on their arguments has important consequences in QCD.  It can
provide a mechanism for quark confinement and is the reason why bulk
thermodynamic quantities, such as the pressure and entropy, approach their
ultrarelativistic limits slowly.

\subsection{Chiral phase transition at $\mu=0$}
\label{secmuzero}
One order parameter for the chiral symmetry restoration transition is the
quark condensate, defined via the renormalised dressed-quark propagator: 
\begin{equation} 
-\,\langle \bar q q \rangle_\zeta^0  =  \lim_{\bar\Lambda\to \infty}
        Z_4(\zeta^2,\bar\Lambda^2)\, N_c \int^{\bar\Lambda}_{l,q}\,{\rm
        tr}_{\rm Dirac} \left[ S_{\hat m  =0}(q) \right]\,.
\end{equation}
Since ${\rm tr}_{\rm Dirac} \left[ S_{\hat m =0}(q) \right]\propto
B(\vec{p},\tilde\omega_k)$, a simpler and equivalent order parameter is
\begin{equation}
\label{chiorder}
{\cal X}(t,h) := {\sf Re}\,B_0(\vec{p}=0,\tilde \omega_0)\,;\;
t:= \frac{T}{T_c} - 1\,,\; h:= \frac{m^\zeta}{T}\,,
\end{equation}
which makes it clear that the zeroth Matsubara mode determines the character
of the chiral phase transition.  (An order parameter for confinement, valid
for both light- and heavy-quarks, was introduced in Ref.$\,$\cite{prl}.  It is
a single, quantitative measure of whether or not a Schwinger function has a
Lehmann representation, and has been used$\,$\cite{m95} to striking effect in
QED$_3$.)

\begin{figure}[t]

\hspace*{0.0cm}\epsfig{figure=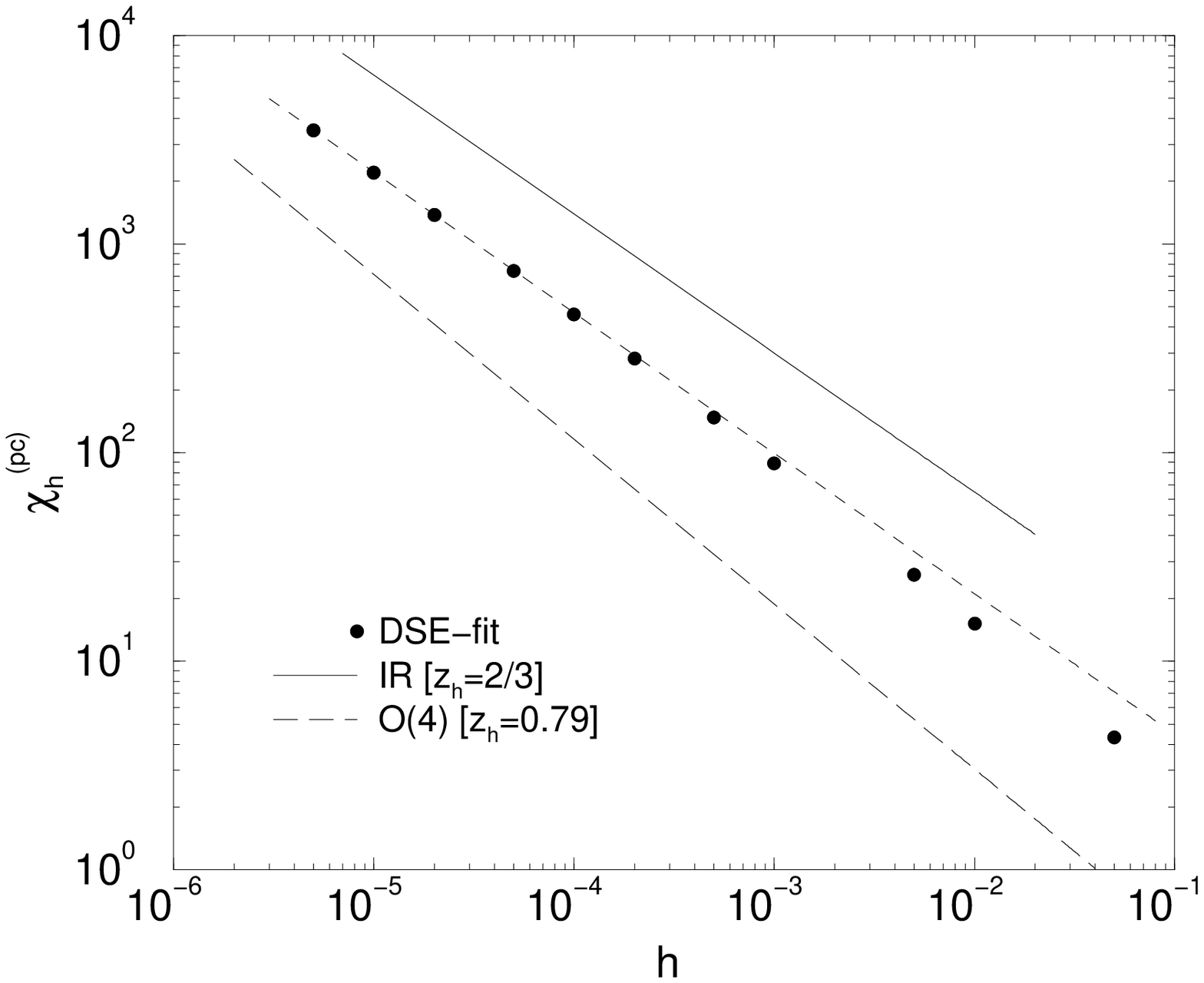,height=6.7cm}
\vspace*{-6.75cm}

\hspace*{7.9cm}\epsfig{figure=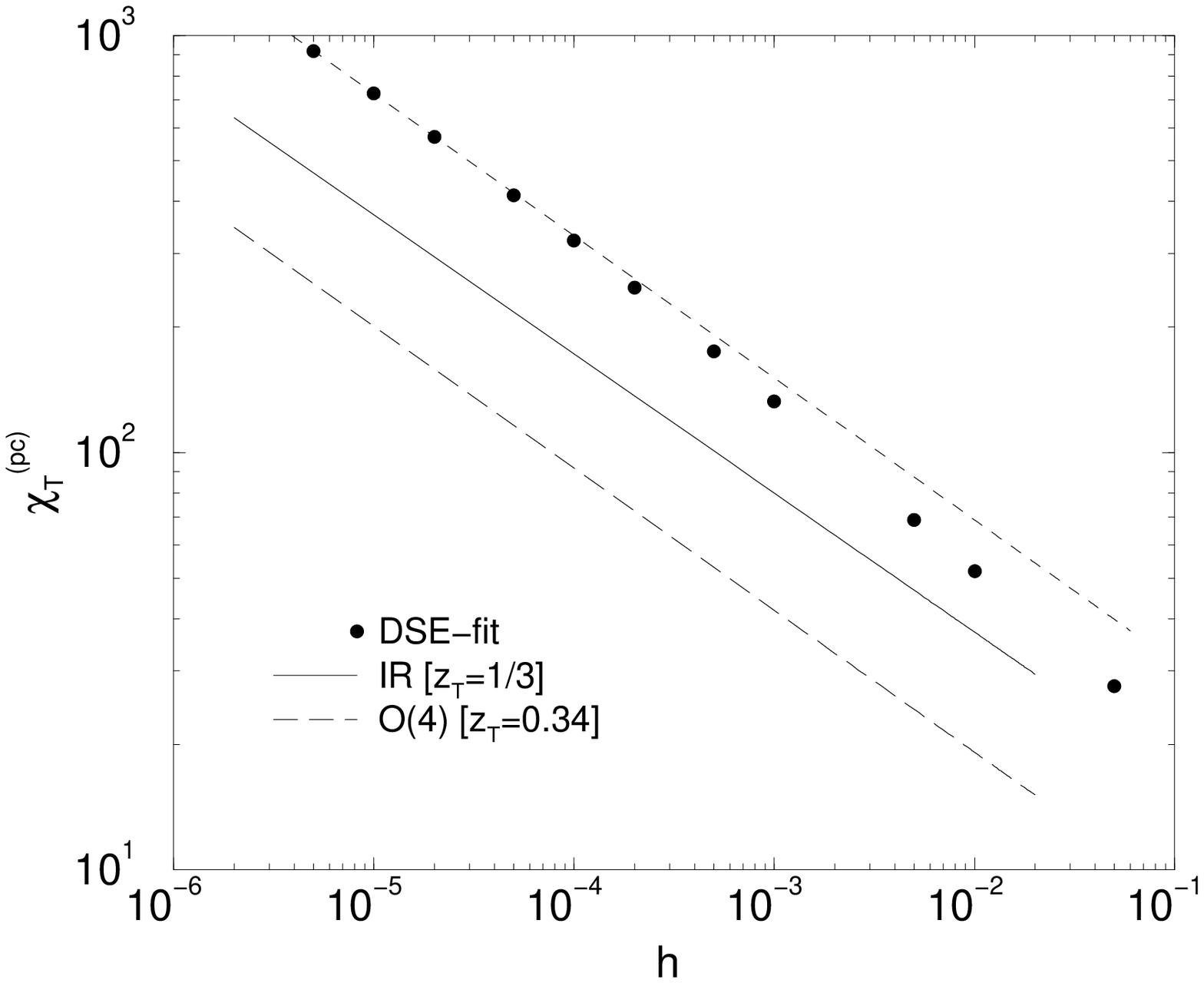,height=6.7cm}\vspace*{-3.0em}
\caption{Behaviour of the susceptibilities in the DSE model of
two-light-flavour QCD illustrated in
Fig.$\,$\protect\ref{gluonpic},\protect\cite{arne} which is a minor modification
of that in Ref.$\,$\protect\cite{mr97}: $\omega \to 0.6\,$GeV, that provides a
slightly better description of $\pi$ and $K$ properties.  The critical
temperature is $T_c^{\cal X} = 152\,$MeV and the critical exponents take
mean-field values: $z_h=0.67$, $z_t= 0.33$; i.e., $\beta=0.5$, $\delta=3.0$,
as might be anticipated because at long-range the interaction $\sim {\rm
const.}$ in configuration space.  The other curves illustrate the slopes
characterising mean-field (labelled IR) and O(4) critical exponents.
\label{suscs}}\vspace*{-1.0em}
\end{figure}

The chiral transition is completely characterised by two critical exponents:
$\beta$, $\delta$, which can be extracted from the behaviour of the chiral
and thermal susceptibilities:$\,$\cite{arneI}
\begin{eqnarray}
\chi_h(t,h) := 
\left.\frac{\partial\, {\cal X}(t,h)}
        {\!\!\!\!\!\!\partial h}\right|_{t} \,,
&&
\chi_t(t,h) := 
\left.\frac{\partial\, {\cal X}(t,h)}
        {\!\!\!\!\!\!\partial t}\right|_{h}\,.
\end{eqnarray}
As a function of $T$ the susceptibilities have a peak, which defines the
pseudocritical points: $t_{\rm pc}^h$, $t_{\rm pc}^t$, and because the
correlation length is infinite at a second order transition it follows that
\begin{eqnarray}
\label{scaling}
\chi_h^{\rm pc}  :=  \chi_h(t_{\rm pc}^h,h) \propto h^{-z_h}\,,\;
        z_h:= 1 - \case{1}{\delta} \,,
&&
\chi_t^{\rm pc}  :=  \chi_t(t_{\rm pc}^t,h) 
        \propto h^{-z_t}\,,
        \;z_t:= \case{1}{\beta\delta}\,(1-\beta)\,.
\end{eqnarray}
Hence the critical exponents can be obtained by plotting $\log_{10}\chi^{\rm
pc}$ vs. $\log_{10}h$, as depicted in Fig.$\,$\ref{suscs}.
This figure makes one thing abundantly clear: very small values of the
current-quark mass must be used to obtain accurate values of the critical
exponents.  If one retains only those values $5\times 10^{-4} \lsim h \lsim
5\times 10^{-2}$ then an apparently good linear fit yields: $z_h=0.78$, $z_t=
0.40$, or $\beta=0.36$, $\delta=4.5$, which are quite close to the values of
the $O(4)$ model.  The small values of the current-quark mass we require to
accurately estimate the critical exponents are inaccessible in contemporary
lattice simulations.

\subsection{$T$ and $\mu$ nonzero}
This is a difficult problem and the most complete study$\,$\cite{thermo} to
date employs a simple {\it Ansatz} for the dressed-gluon propagator that
exhibits only the infrared enhancement suggested by Ref.$\,$\cite{bp89}:
\begin{equation}
\label{mnprop}
g^2 D_{\mu\nu}(\vec{p},\Omega_k) = 
\left(\delta_{\mu\nu} 
- \frac{p_\mu p_\nu}{|\vec{p}|^2+ \Omega_k^2} \right)
2 \pi^3 \,\frac{\eta^2}{T}\, \delta_{k0}\, \delta^3(\vec{p})\,,
\end{equation}
with $\Omega_k=2 k \pi T$ and $\eta$ a mass-scale.  It is an
infrared-dominant model and does not represent well the behaviour of
$D_{\mu\nu}(\vec{p},\Omega_k)$ away from \mbox{$|\vec{p}|^2+ \Omega_k^2
\approx 0$}.  Consequently some model-dependent artefacts arise.  However,
there is significant merit in its simplicity and, since the artefacts are
easily identified, the model remains useful as a means of elucidating many of
the qualitative features of more sophisticated {\it Ans\"atze}.

Using Eq.$\,$(\ref{mnprop}) and the rainbow truncation [$\Gamma_\mu \to
\gamma_\mu$ in Eq.$\,$(\ref{regself})], the quark DSE is$\,$\cite{bender96}
\begin{equation}
\label{mndse}
S^{-1}(\vec{p},\omega_k) = S_0^{-1}(\vec{p},\tilde \omega_k)
        + \case{1}{4}\eta^2\gamma_\nu S(\vec{p},\tilde \omega_k) \gamma_\nu\,,
\end{equation}
and we see that the simplicity of the {\it Ansatz} allows the reduction of an
integral equation to an algebraic equation.  Its solution exhibits many of
the qualitative features of more sophisticated models.

In the chiral limit Eq.$\,$(\ref{mndse}) has two qualitatively distinct
solutions.  The Nambu-Goldstone solution, with
\begin{equation}
\label{ngsoln}
\!\!
\begin{array}{ll}
B(\tilde p_k)  = \left\{
\begin{array}{ll}
\sqrt{\eta^2 - 4 \tilde p_k^2}\,,
        & \!\! {\sf Re}(\tilde p_k^2)<\case{\eta^2}{4}\\
    0\,,&  \!\!{\rm otherwise}
\end{array}\right.      \!\!\!\! ,
&
\!\!
C(\tilde p_k)  = \left\{
\begin{array}{ll}
2\,, &  \!\!{\sf Re}(\tilde p_k^2)<\case{\eta^2}{4}\\
\case{1}{2}\left( 1 + \sqrt{1 + \case{2 \eta^2}{\tilde p_k^2}}\right)
\,,&  \!\!{\rm otherwise}
\end{array}\right. \!\!\!\! \! 
\end{array},
\end{equation}
where $\tilde p_l := (\vec{p},\tilde\omega_l)$, describes a phase of the
model in which: 1) chiral symmetry is dynamically broken, because one has a
nonzero quark mass-function, $B(\tilde p_k)$, in the absence of a
current-quark mass; and 2) the dressed-quarks are confined, because the
propagator described by these functions does not have a Lehmann
representation.  The alternative Wigner solution, for which
\begin{eqnarray}
\label{wsoln}
\hat B(\tilde p_k)  \equiv  0 &,\;& 
\hat C(\tilde p_k)  = 
\case{1}{2}\left( 1 + \sqrt{1 + \case{2 \eta^2}{\tilde p_k^2}}\right)\,,
\end{eqnarray}
describes a phase of the model with neither DCSB nor confinement.  

The relative stability of the different phases is measured by a
$(T,\mu)$-dependent vacuum pressure difference: ${\cal B}(T,\mu)$.  ${\cal
B}(T,\mu)=0$ defines the phase boundary, and the deconfinement and chiral
symmetry restoration transitions are coincident.  For $\mu=0$ the transition
is second order and the critical temperature is $T_c^0 = 0.159\,\eta$, which
using the value of $\eta=1.06\,$GeV obtained by fitting the $\pi$ and $\rho$
masses corresponds to $T_c^0 = 0.170\,$GeV.  This is only 12\% larger than
the value reported in Sec.$\,$\ref{secmuzero}, and the order of the transition
is the same.  For any $\mu \neq 0$ the transition is first-order.  For $T=0$
the critical chemical potential is $\mu_c^0=0.3\,$GeV, which is
\mbox{$\approx 30$\%} smaller than the result in Ref.$\,$\cite{greg}.  The
discontinuity in the order parameter vanishes at $\mu=0$.

\begin{figure}[t]
\vspace*{-1.0em}

\centering{\ \epsfig{figure=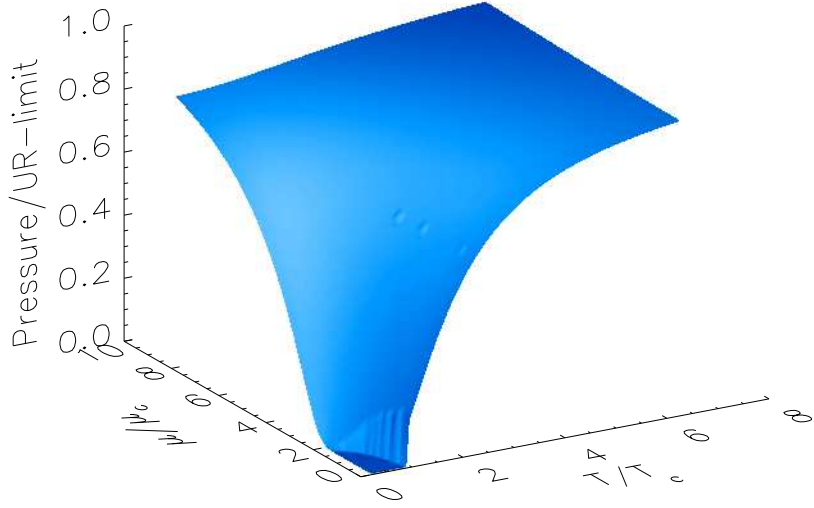,height=7.5cm} }\vspace*{-1.0em}
\caption{The quark pressure, $P_q(\bar T,\bar\mu)$, normalised to the free,
massless (or Ultra-Relativistic) result.\label{presfig}}\vspace*{-1.0em}
\end{figure}

The quark pressure is easily calculated and is depicted in
Fig.$\,$\ref{presfig}.  Confinement means that $P_q \equiv 0$ in the confined
domain.  In the deconfined domain it approaches the ultrarelativistic, free
particle limit, $P_{\rm UR}$, at large values of $ T$ and $\mu$ but the
approach is slow.  For example, at $ T \sim 2 T_c^0$, or $ \mu \sim 3
\mu_c^0$, $P_q$ is only $0.5\,P_{\rm UR}$.  This feature results from the
persistence of momentum dependent modifications of the quark propagator into
the deconfined domain, as evident with $C\not\equiv 1$ in Eq.$\,$(\ref{wsoln}).
The figure also highlights the ``mirroring'' of finite-$T$ behaviour in the
$\mu$-dependence of the bulk thermodynamic quantities.

\begin{figure}[t]

\hspace*{-0.30cm}\epsfig{figure=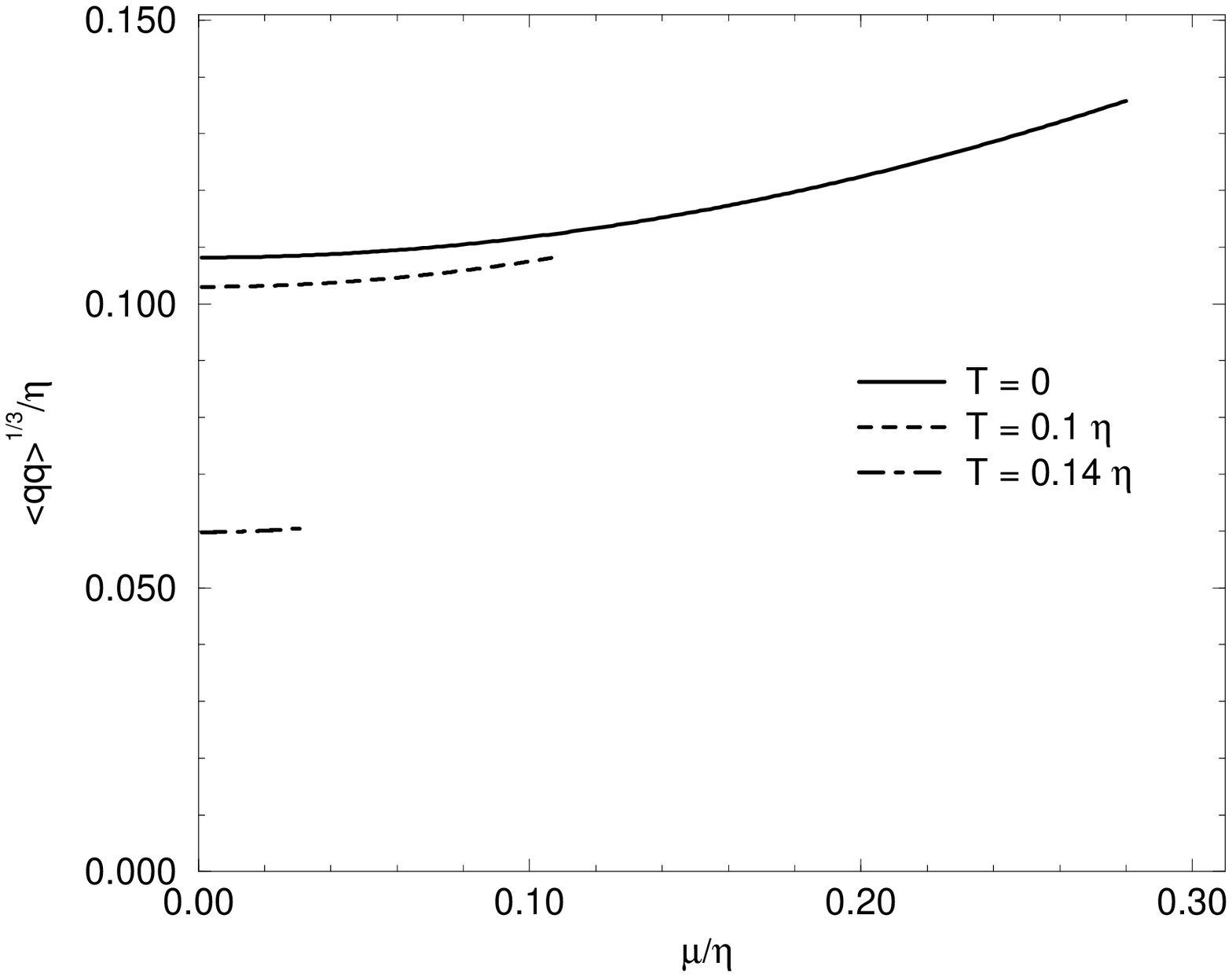,height=7.5cm}
\vspace*{-7.53cm}

\hspace*{7.6cm}\epsfig{figure=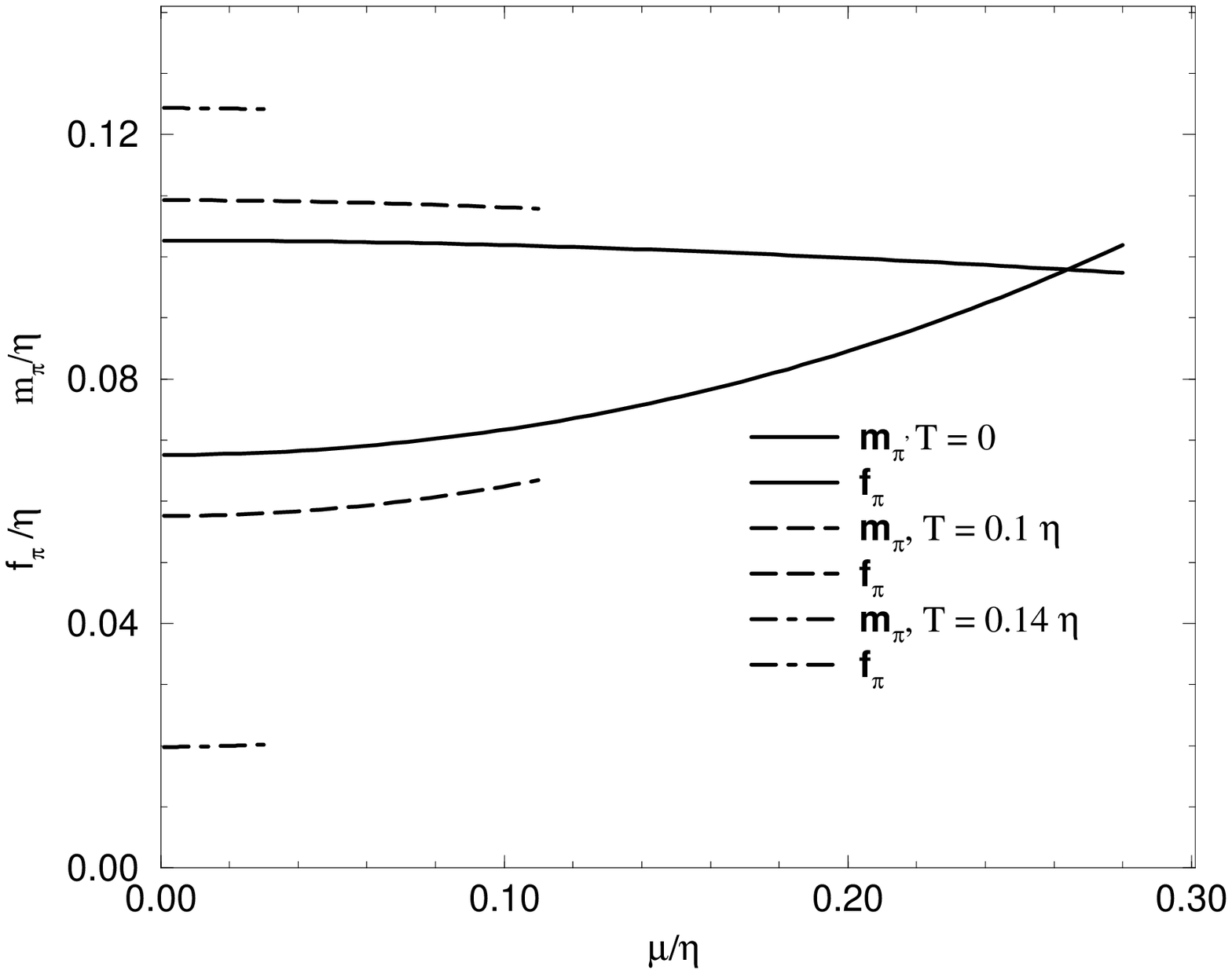,height=7.5cm}\vspace*{-4.0em}
\caption{$-\langle \bar q q \rangle$, $f_\pi$ and $m_\pi$ as a function of
$\mu$ for a range of values of $T$.  $m_\pi$ is only weakly sensitive to
changes in $\mu$ and $T$ because of mutual cancellations between the effects
of these intensive variables on $-\langle \bar q q \rangle$ and $f_\pi$.
That is a result of Goldstone's theorem; i.e., the preservation of the
axial-vector Ward-Takahashi identity.\label{piobs}}\vspace*{-1.0em}
\end{figure}

The vacuum quark condensate is given by the simple
expression$\,$\cite{schmidt98}
\begin{equation}
\label{qbq}
-\langle \bar q q \rangle = 
\eta^3\,\frac{8 N_c}{\pi^2} \bar T\,\sum_{l=0}^{l_{\rm max}}\,
\int_0^{\bar\Lambda_l}\,dy\, y^2\,
{\sf Re}\left( \sqrt{\case{1}{4}- y^2   
+ \bar \mu^2 - \bar \omega_{l}^2 - 2 \,i\,\bar \mu \,\bar \omega_l }\right)\,:
\end{equation}
$\bar T=T/\eta$, $\bar \mu=\mu/\eta$; $l_{max}$ is the largest value of $l$
for which $\bar\omega^2_{l_{\rm max}}\leq (1/4)+\bar\mu^2$ and this also
specifies $\omega_{l_{max}}$, $\bar\Lambda^2 = \bar\omega^2_{l_{\rm
max}}-\bar\omega_l^2$, $\bar p_l = (\vec{y},\bar\omega_l+i\bar\mu)$.  At
$T=0=\mu$, $(-\langle \bar q q \rangle) = \eta^3 /(80\,\pi^2) = (0.11\,
\eta)^3$.  Obvious from Eq.$\,$(\ref{qbq}) is that $(-\langle \bar q q \rangle)$
decreases continuously to zero with $T$ but {\it increases} with $\mu$, up to
a critical value of $\mu_c(T)$ when it drops discontinuously to zero.  This
behaviour is driven by the combination $\mu^2 - \omega_l^2$ and is exhibited
by the calculated result depicted in Fig.$\,$\ref{piobs}.
In the chiral limit one also has the simple expression
\begin{eqnarray}
\label{npialg}
f_\pi^2 & = & \eta^2 \frac{16 N_c }{\pi^2} \bar T\,\sum_{l=0}^{l_{\rm max}}\,
\frac{\bar\Lambda_l^3}{3} \left( 1 + 4 \,\bar\mu^2 - 4 \,\bar\omega_l^2 -
\case{8}{5}\,\bar\Lambda_l^2 \right)\,.
\end{eqnarray}
It too involves the combination $\mu^2 - \omega_l^2$ and without calculation
Eq.$\,$(\ref{npialg}) indicates that $f_\pi$ will {\it decrease} with $T$ and
{\it increase} with $\mu$, as exhibited by calculated result depicted in
Fig.$\,$\ref{piobs}.  These results confirm those obtained$\,$\cite{prl,greg}
with more sophisticated {\it Ans\"atze}, and in doing so provide for their
algebraic elucidation: the behaviour is a consequence of the necessary
momentum dependence of the dressed-quark self energy.

\begin{figure}
\vspace*{-1.0em}

\centering{\
\epsfig{figure=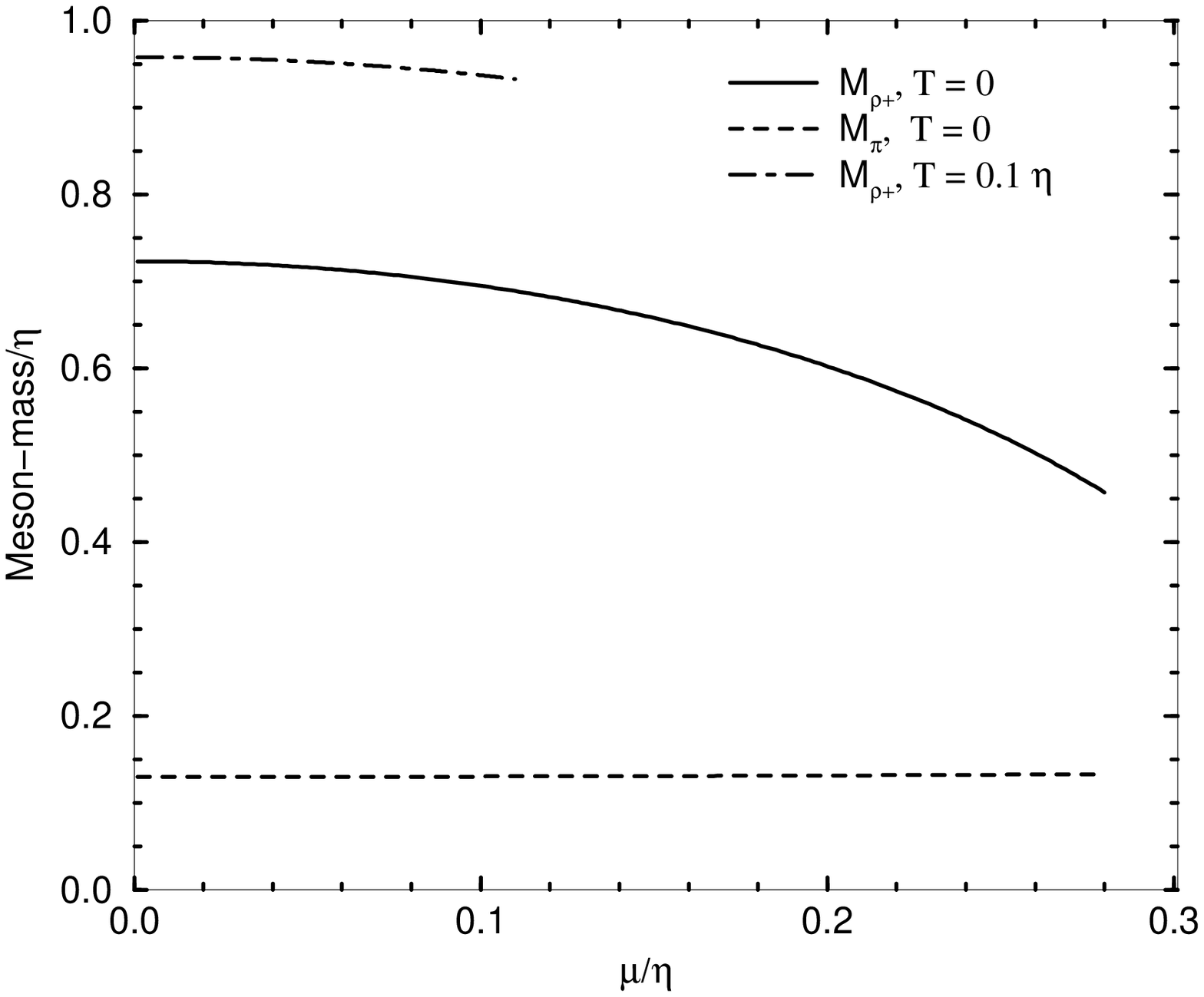,height=7.90cm}}\vspace*{-3.0em}
\caption{\label{pirhomass} $M_{\rho}=M_{\omega}$ and $m_\pi$ as a function
of $\bar\mu$ for $\bar T = 0, 0.1$.  On the scale of this figure, $m_\pi$ is
insensitive to this variation of $T$.  The current-quark mass is $m=
0.011\,\eta$, which for $\eta=1.06\,$GeV yields $M_{\rho}= 770\,$MeV and
$m_\pi=140\,$MeV at $T=0=\mu$.}\vspace*{-1.0em}
\end{figure}

The $(T,\mu)$-response of meson masses is determined by the ladder BSE
\begin{equation}
\label{bse}
\Gamma_M(\tilde p_k;\check P_\ell)= - \frac{\eta^2}{4}\,
{\sf Re}\left\{\gamma_\mu\,
S(\tilde p_i +\case{1}{2} \check P_\ell)\,
\Gamma_M(\tilde p_i;\check P_\ell)\,
S(\tilde p_i -\case{1}{2} \check P_\ell)\,\gamma_\mu\right\}\,,
\end{equation}
where $\check P_\ell := (\vec{P},\Omega_\ell)$, with the bound state mass
obtained by considering $\check P_{\ell=0}$.  In this truncation the
$\omega$- and $\rho$-mesons are degenerate.  The calculated mass of the
$\pi$- and $\rho$-mesons is depicted in Fig.$\,$\ref{pirhomass}.
The behaviour is easily understood by again considering the chiral limit
where the mass of the longitudinal component of the $\rho$-meson is 
\begin{equation}
\label{mplus}
M_{\rho}^2 = \case{1}{2} \eta^2 - 4 (\mu^2 - \pi^2 T^2)\,.
\end{equation}
The characteristic combination $\mu^2 - \pi^2 T^2$ appears again.  It is
responsible for the anticorrelation between the response of $M_{\rho}$ to
$T$ and its response to $\mu$, and that $M_{\rho}^2$ rises linearly with
$T^2$ for $\mu=0$, just like a gauge-boson Debye mass.  The mass of the
transverse component of the vector meson is insensitive to $T$ and $\mu$.

In a 2-flavour, free-quark gas at $T=0$ the baryon number density is $\rho_B=
2 \mu^3/(3 \pi^2)\,$, by which gauge nuclear matter density,
$\rho_0=0.16\,$fm$^{-3}$, corresponds to $\mu= \mu_0 := 260\,$MeV$\,=
0.245\,\eta$.  At this chemical potential the algebraic model yields
\begin{eqnarray}
\label{mrhoa}
M_{\rho}(\mu_0)  \approx  0.75 M_{\rho}(\mu=0) &,\; &
M_{\phi}(\mu_0)  \approx  0.85 M_{\phi}(\mu=0)\,,
\end{eqnarray}
where $M_{\phi}(\mu=0)=1.02\,$GeV for $m_s=180\,$MeV.  The study of
Ref.$\,$\cite{greg} indicates that a more realistic representation of the
ultraviolet behaviour of the dressed-gluon propagator expands the horizontal
scale in Fig.$\,$\ref{pirhomass}, with the critical chemical potential
increased by 25\%.  This suggests that a better estimate is obtained by
evaluating the mass at $\mu_0^\prime=0.20\,\eta$, which yields
\begin{eqnarray}
\label{mrhob}
M_{\rho}(\mu_0^\prime) \approx  0.85 M_{\rho}(\mu=0) &,\; &
M_{\phi}(\mu_0^\prime) \approx  0.90 M_{\phi}(\mu=0) \,;
\end{eqnarray}
a small, quantitative modification.  The difference between
Eqs.$\,$(\ref{mrhoa}) and (\ref{mrhob}) is a measure of the theoretical
uncertainty in the estimates in each case.  Pursuing this suggestion further,
$\mu_2=\mu_0^\prime\,^3\!\!\!\surd 2$, corresponds to $2\rho_0$, at which
point $M_{\omega}= M_{\rho} \approx 0.72\, M_{\rho}(\mu=0)$ and $M_{\phi}
\approx 0.85\, M_{\phi}(\mu=0)$, while at the $T=0$ critical chemical
potential, which corresponds to approximately $3\rho_0$ in Ref.$\,$\cite{greg},
$M_{\omega}= M_{\rho} \approx 0.65\, M_{\rho}(\mu=0)$ and $M_{\phi}
\approx 0.80\, M_{\phi}(\mu=0)$.  These are the maximum possible reductions
in the meson masses.

This simple model preserves the momentum-dependence of gluon and quark
dressing, which is an important qualitative feature of more sophisticated
studies.  Its simplicity means that many of the consequences of that dressing
can be understood algebraically.  For example, it elucidates the origin of an
anticorrelation, found for a range of quantities, between their response to
increasing $T$ and that to increasing $\mu$.  That makes clear why the
transition to a QGP is second order with increasing $T$ and first order with
$\mu$.  Further it provides an algebraic explanation of why the
$(T,\mu)$-dependence of $(-\langle \bar q q)\rangle$ and $f_\pi$ must be {\it
opposite} to that of $M_{\rho}$.

\section{DIQUARKS}
It is not implausible that diquark (quark-quark) correlations play an
important role in the strong interaction.  To see why, consider the class of
models that can be characterised by an effective interaction of the form:
\begin{equation}
\label{stupid}
\int \,d^4x\,d^4y\,\bar q(x) \gamma_\mu\frac{\lambda^a}{2} q(x)\,g^2
                D_{\mu\nu}(x-y)\, \bar q(y) \gamma_\nu \frac{\lambda^a}{2}
                q(y)\,.
\end{equation}
Field theories defined by such an interaction admit$\,$\cite{cahill} a
meson-diquark bosonisation that at tree-level predicts a mass for both mesons
and diquarks.  The procedure corresponds closely to solving the
Bethe-Salpeter equations obtained by combining the ladder truncation of the
quark-antiquark and quark-quark scattering kernels, depicted in
Fig.$\,$\ref{skeleton}, with the rainbow truncation of the quark DSE.  Any
form of $g^2 D_{\mu\nu}(x-y)$ that is accurately able to reproduce the mass
of the $\pi$ and $\rho$ mesons in this truncation will
predict$\,$\cite{luqian} the existence of stable, colour-antitriplet diquark
bound states with masses:
\begin{equation}
m_{0^+}^{ud} \approx 740\,{\rm MeV},\; 
m_{1^+}^{ud} \approx 950\,{\rm MeV},\;
m_{1^-}^{ud} \approx m_{0^-}^{ud} \approx 1500\,{\rm MeV}.
\end{equation}
This points to a possibly important attraction in the quark-quark channel.

However, it also presents a problem, of course, because asymptotic diquark
states are not observed.  The defect is in the truncation.$\,$\cite{bender96}
If one proceeds beyond the ladder-truncation in the quark-quark scattering
kernel; e.g., including the other diagrams depicted in
Fig.$\,$\ref{skeleton}, one uncovers a repulsive contribution from the
crossed-box diagram that eliminates the pole in the quark-quark scattering
matrix; i.e., the full kernel does not support spurious diquark bound states.
(The same procedure applied to SU$(N_c=2)$-QCD predicts$\,$\cite{qcII} that
the mesons and diquarks are degenerate in that theory.)  It is clear
therefore that analyses of diquark correlations based solely on an effective
interaction of the type in Eq.$\,$(\ref{stupid}) can yield erroneous results,
and the suggestion of diquark condensation may be amongst them.

The consequences of a putative diquark condensate are easy to elucidate.  It
is adequate to consider a simple separable model$\,$\cite{sep}
\begin{equation}
g^2 D_{\mu\nu}(\vec{p}-\vec{q},\omega_k-\omega_l)
:= \delta_{\mu\nu} \, D_0\,g(\vec{p}\,^2+\omega_k^2)
                        \,g(\vec{q}\,^2+\omega_l^2)\,,\;
g(k^2):= \exp(-k^2/\Lambda^2)\,,
\end{equation}
which preserves Goldstone's theorem and ensures the absence of a Lehmann
representation for the quark propagator.  This {\it Ansatz} has two
parameters that can be fixed by requiring a good description of pion
properties at $T=0=\mu$; e.g., $D_0 = 106/\Lambda^2$, $\Lambda=0.71\,$GeV
yield $f_\pi = 93\,$MeV and $-\langle \bar q q\rangle = (238\,{\rm MeV})^3$.
A feature of the model is that it admits a semi-algebraic analysis and at
$\mu=0$, with these parameter values, it has second-order deconfinement and
chiral symmetry restoration transitions at $T_c\gsim 130\,$MeV, while for
$T=0$ these transitions are first-order at $\mu_c = 330\,$MeV neglecting the
effect of a diquark condensate.

\begin{figure}
\vspace*{-1.0em}

\centering{\
\epsfig{figure=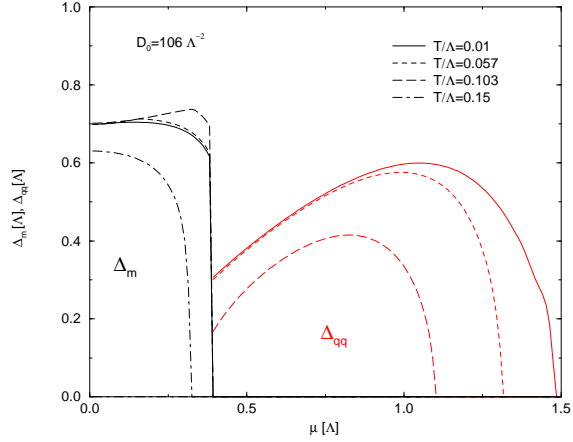,height=7.5cm,angle=-90}}\vspace*{-2.0em}
\caption{\label{diquarks} Quark and diquark mass gaps: $\Delta_m$,
$\Delta_{qq}$. $\Delta_{qq}\equiv 0$ at $T=0.15\Lambda$,
$\Lambda=0.71\,$GeV.}\vspace*{-1.0em}
\end{figure}

A meson-diquark bosonisation yields a tree-level auxiliary-field effective
action whose extremum is determined by a pair of coupled equations for the
quark and diquark mass gaps, $\Delta_m$ and $\Delta_{qq}$:
\begin{eqnarray}
\Delta_m & = & \Delta_m \,\case{8}{3}D_0 \,
T\sum_{l=-\infty}^\infty\int\frac{d^3p}{(2\pi)^3}\,
\frac{g^2(\tilde p_l^2)}{\epsilon_q(\tilde p_l^2)}\,
\left[ \frac{\epsilon_q(\tilde p_l^2)+\mu}
                {\omega_l^2 + (\epsilon_q(\tilde p_l^2)+\mu)^2 
        + \epsilon_{qq}^2(\tilde p_l^2) }
        + \left\{\mu \to -\mu \right\}       \right]\!, \\
\Delta_{qq}  & = &  \Delta_{qq}\,\case{4}{3}D_0 \,
T\sum_{l=-\infty}^\infty\int\frac{d^3p}{(2\pi)^3}\,
        g^2(\tilde p_l^2)
        \left[ \frac{1}{\omega_l^2 + (\epsilon_q(\tilde p_l^2)+\mu)^2 
        + \epsilon_{qq}^2(\tilde p_l^2) }
        + \left\{\mu \to -\mu \right\}         \right]\!,
\end{eqnarray}
with
\begin{equation}
\epsilon_q^2(\tilde p_l^2) = \vec{p}\,^2 + \Delta_m^2 \, g^2(\tilde p_l^2)\,,\; 
\epsilon_{qq}^2(\tilde p_l^2)= \Delta_{qq}^2\, g^2(\tilde p_l^2 )\,.
\end{equation}
The first equation arises from the confined-quark contribution to the
effective action, while the second is the contribution from unconfined
diquarks.  The preliminary result of Ref.~\cite{sep} for the
$(T,\mu)$-dependence of $\Delta_m$ and $\Delta_{qq}$ is depicted in
Fig.$\,$\ref{diquarks}.
The indications are that at $T=0$ there is a first order transition from a
$\Delta_m \neq 0$ phase to a $\Delta_{qq} \neq 0$ phase at $\mu\approx
280\,$MeV.  This behaviour persists until $T\simeq 100\,$MeV when only
$\Delta_m \neq 0$ is possible.  There appears to be a tricritical point in
the $(T,\mu)$-plane.

Clearly the formation of a diquark condensate is a simple and direct
consequence of assuming that attraction dominates in the quark-quark
scattering matrix.  However, equally clearly, the validity of that assumption
requires further consideration, with a first subject being the effect of the
often-ignored repulsive terms in the quark-quark scattering kernel.

\section{CLOSING REMARKS}
The DSEs are an efficacious tool for studying the strong interaction.  In
this application they expose the qualitative importance and quantitative
effect of the nonperturbative dressing of propagators and vertices in QCD.
The modelling involved is based on this observation, and it is necessary
because of the need to make truncations.  Questions will always be asked
about the fidelity of that modelling, and the answers will come as more is
learnt about about the constraints that Ward and Slavnov-Taylor identities in
the theory can provide.  That approach has already been particularly fruitful
in QED,\cite{ayse97} and in the development of a systematic truncation
procedure for the kernel of the quark DSE and meson BSE.\cite{bender96,qcII}
In the meantime the judicious application of DSEs will continue to provide a
flexible, intuition-building framework for the prediction, correlation and
validation of observables.

%eeeeeeeeeeeeeeeeeeeeeeeeeeeeeeeeeeeeeeee
\end{document}